\documentclass[12pt]{iopart}
\pdfoutput=1
\expandafter\let\csname equation*\endcsname\relax
\expandafter\let\csname endequation*\endcsname\relax
\usepackage{mathtools} 
\usepackage[utf8x]{inputenc}
\usepackage{epsfig}
\usepackage{amsmath} 
\usepackage{amssymb}
\usepackage{color}
\usepackage[toc,page]{appendix}
\usepackage{float}
\usepackage{graphicx}
\usepackage{anysize}
\usepackage{xfrac}
\usepackage{soul}
\soulregister\cite7
\soulregister\ref7
\soulregister\pageref7

\usepackage{soul}

\marginsize{3cm}{3cm}{2.5cm}{4.5cm}
\begin{document}

\title[Anomalous mobility of a driven active particle in a steady laminar flow]{Anomalous mobility of a driven active particle in a steady laminar flow}

\author{F. Cecconi, A. Puglisi, A. Sarracino}
\address{CNR-ISC and Dipartimento di Fisica, Sapienza Universit\`a di Roma, p.le A. Moro 2, 00185 Roma, Italy}
\ead{fabio.cecconi@roma1.infn.it,andrea.puglisi@roma1.infn.it, alessandro.sarracino@roma1.infn.it}
\author{A. Vulpiani}
\address{Dipartimento di Fisica, Sapienza Universit\`a di Roma and CNR-ISC, p.le A. Moro 2, 00185 Roma, Italy}
\address{Centro Interdisciplinare B. Segre, Accademia dei Lincei, Roma, Italy}
\ead{angelo.vulpiani@roma1.infn.it}
\vspace{10pt}

\begin{abstract}
We study, via extensive numerical simulations, the force-velocity
curve of an active particle advected by a steady laminar flow, in the
nonlinear response regime. Our model for an active particle relies on
a colored noise term that mimics its persistent motion over a time
scale $\tau_A$.  We find that the active particle dynamics shows
non-trivial effects, such as negative differential and absolute
mobility (NDM and ANM, respectively). We explore the space of the
model parameters and compare the observed behaviors with those
obtained for a passive particle ($\tau_A=0$) advected by the same
laminar flow. Our results show that the phenomena of NDM and ANM are
quite robust with respect to the details of the considered noise: in
particular for finite $\tau_A$ a more complex force-velocity relation
can be observed.
\end{abstract}

\section{Introduction}
The study of the motion of a tracer driven by an external force in a
complex environment plays a central role in many research areas,
ranging from transport phenomena at different
scales~\cite{Pollutant,CEN13,SBMM13,MEH16,VBMWO17,CGCB17}, to the
response theory in non-equilibrium
systems~\cite{BPRV08,seifertrev,BM13}. Although some results have been
obtained in the last decades in the framework of linear response, the
behavior of the tracer for perturbations in the nonlinear regime is
much less understood. It can show complex behaviors, featuring
surprising phenomena such as negative differential mobility (NDM) and
even absolute negative mobility (ANM). The former means a
non-monotonic behavior of the force-velocity curve, while the latter
indicates that, for certain values of the model parameters, the tracer
can display a stationary velocity opposite to the direction of the
applied external force. NDM has been recently studied in discrete
lattice gas models, where also some analytical results are
available~\cite{LF13,BM14,BIOSV14,BIOSV16,CMP18}, while ANM is
observed in systems with continuous states as reported for instance
in~\cite{RERDRA05,MKTLH07,ERAR10,SCPV16,CPSV17}.  The general
mechanisms responsible for these phenomena rely on some trapping
effects occurring in the system due to the coupling of the tracer
dynamics with the surrounding complex environment, and depend on the
specific model.

In this paper we study a similar problem for the case of an active
particle advected by a steady laminar flow. Active matter is generally
characterized by self propulsion, namely an internal conversion of
energy into unidirected motion, which results in a persistent direction,
over an ``active'' timescale $\tau_A$.  Instances of such systems range
from biological organisms~\cite{ram10,cavagna12} to man-made
devices~\cite{dileo16}.  Several models have been proposed to study
the dynamics of these non-equilibrium systems, that can show
interesting phenomena, such as phase separation, clustering, symmetry
breaking and so on~\cite{ram10,dauchot13,cates15}.  The response of
active particles moving in complex environments under the action of an
external force, is an important issue in different contexts: on the
one hand, from a theoretical perspective, this problem plays a central
role in the theory of non-equilibrium (also in the non-linear regime)
fluctuation-dissipation
relations~\cite{LCSZ08a,LCSZ08b,basu,szamel,helden}; on the other
hand, the study of the driven dynamics of active particles has
important applications in active microrheology~\cite{puertas} as well
as in the modeling of the dispersion of micro-organisms in
fluids~\cite{GRS12,SDMB16}.  Due to the non-equilibrium
nature of these systems, non-trivial behaviors such as NDM are
expected, as observed in recent molecular dynamics simulations of
active matter particles moving through a random obstacle
array~\cite{reich}, or in discrete models on a
lattice~\cite{CP13,BZBTV17}. Many other interesting effects can be
observed due to the coupling of the active (non-equilibrium) motion of
the particles with the boundaries~\cite{RR07,MBGL15,DGCHS15,MSMP17} or
with the surrounding crowding
environment~\cite{SS08,MMO11,HF13,AT16,PAG17,MCCB17}.  In particular,
driven motion along narrow channels or confined geometries can show
subdiffusion and anomalous
fluctuations~\cite{NMP07,BHM09,BBCI13,FBCV13,BCCC13,BIOSV15,BT15,CMP17}.

Here we consider a different approach to describe active matter, where
the persistent motion is introduced via a colored noise term, with a
finite correlation time $\tau_A$ \cite{hanggi95,MBGL15}.  Inspired by
this model, we study the nonlinear response to an external bias of an
active particle advected by a divergenceless velocity field in the
presence of colored noise. This model generalizes the system considered
in~\cite{SCPV16,CPSV17}, where the dynamics of an inertial particle
was considered.  Indeed, in the limit of $\tau_A\to 0$ one recovers
the $\delta-$correlated ``passive'' case, while for $\tau_A\to \infty$
one obtains the deterministic, zero-noise dynamics.  As shown in
previous works~\cite{SCPV16,CPSV17}, the effective motion of the
tracer in the passive case, for small noise, occurs along preferential
``channels'', that can be aligned downstream or upstream with respect
to the force, resulting in a non-trivial force-velocity relation.
Here we show that phenomena such as NDM and ANM also take place for
the active particle dynamics and therefore, in these models, they are
rather robust with respect to the kind of considered noise.  Moreover,
we investigate some regions of the parameter space of the model,
identifying the cases where NDM or ANM occur. In particular, we focus
on the range of parameters where in the passive case with $\tau_A=0$
ANM occurs~\cite{CPSV17}: our results clarifies in what manner the
behaviors for $\tau_A=0$ get modified for finite $\tau_A$, and show
that the effect of ANM can be amplified by taking larger values of
$\tau_A$. Indeed, both the range of forces where ANM is observed and
the negative values of the stationary velocity, can be increased by
increasing the persistence time of the particle.

\section{Model}
Consider a driven tracer particle, moving in two dimensions with
position $\mathbf x= (x,y)$ and velocity $\mathbf v =(v_x,v_y)$,
advected by a velocity field $\mathbf U = (U_x,U_y)$. The equations of
motion are the following
\begin{eqnarray}
\dot{x}&=&v_x,  \label{eq0} \\
\dot{y}&=&v_y, \label{eq1} \\ 
\dot{v}_x&=&-\frac{1}{\tau_S}(v_x-U_x)+F +w_x \label{eq2}, \\
\dot{v}_y&=&-\frac{1}{\tau_S}(v_y-U_y)+ w_y, \label{eq2a}
\label{model}
\end{eqnarray}
where $\mathbf U$ is
a divergenceless cellular flow
defined by a stream-function $\psi$ as:
\begin{equation} 
U_x=\frac{\partial \psi(x,y)}{\partial y}, \qquad U_y=-
\frac{\partial \psi(x,y)}{\partial x}\;. 
\label{eq:psi}
\end{equation}
In the above equations $\tau_S$ is the Stokes time, $F$ the external
force in the $x-$direction, and
\begin{equation}
\psi(x,y) = \frac{U_0}{k} \sin(k x)\sin(k y),
\end{equation}
where $k=2\pi/L$.  $w_x$ and $w_y$ are stochastic terms described by
an Ornstein-Uhlenbeck process
\begin{eqnarray}
\dot{w}_x&=&-\frac{w_x}{\tau_A}+\frac{\sqrt{2D_0}}{\tau_A}\xi_x \label{eq3}, \\
\dot{w}_y&=&-\frac{w_y}{\tau_A}+ \frac{\sqrt{2D_0}}{\tau_A}\xi_y,
\label{model2}
\end{eqnarray}
where $\xi_x$ and $\xi_y$ are uncorrelated white noises with zero mean
and variance
\begin{equation}
\langle \xi_\alpha(t)\xi_\beta(t')\rangle =\delta(t-t')\delta_{\alpha\beta}.
\end{equation}
We set $U_0=1$ and $L=1$, and the typical time scale of the flow
becomes $\tau^*=L/U_0=1$. The parameter $\tau_A$ represents the
correlation time of the noise. The limit $\tau_A\to 0$ recovers
the case of uncorrelated noise, and the microscopic thermal noise with
diffusivity $D_0$ can be expressed in terms of the temperature $T$ of
the environment by the relation $D_0=T/\tau_S$.  In the opposite limit
$\tau_A\to \infty$, somehow the system approaches the deterministic
(zero-noise) dynamics, because the stochastic terms $w_x$ and $w_y$
in Eqs.~(\ref{eq2}) and~(\ref{eq2a}) are negligible (order
$\sqrt{D_0/\tau_A}$).  The overdamped version of this model (with no
external force) has been introduced to study the transport of a fluid
particle in the upper mesoscale ocean~\cite{LPVZ95,GOPR95} and has
been analyzed using multiscale technique in~\cite{CC99}.

\section{Anomalous behaviors of the force-velocity curve}
When a tracer particle is driven by a small external force in a simple
(equilibrium) fluid, one expects that the asymptotic mean velocity
selected by the tracer will increase with the applied force, in
agreement with the linear response. However, when the system is out of
equilibrium, due to the presence of currents, or when the applied
force is beyond the linear regime, the behavior of the force-velocity
curve can be highly non-trivial, showing surprising behaviors.

Considering models defined in continuous space, NDM and ANM can be
observed when the dimensionality of the phase space is larger than
two~\cite{ERCB05,SER07}. For instance, ANM can be shown by: i) a
one-dimensional inertial Brownian particle subjected to a periodic
time-dependent force~\cite{MKTLH07}; ii) an overdamped Brownian
particle in two dimensions subjected to dichotomous
noise~\cite{ERH02}; iii) an inertial particle in two dimensions
advected by a velocity field~\cite{SCPV16}. In the first two cases a
detailed analysis of the deterministic properties of the system
identifies the subtle interplay between the stability of coexisting
attractors, noise induced metastability, and transient chaos as the
underlying physical mechanism~\cite{SER07}.  In the latter case iii),
corresponding to the limit $\tau_A=0$ of the model introduced above,
the behavior of the force-velocity curve of the tracer has been
analyzed as a function of the Stokes time $\tau_S$ and of the
amplitude of microscopic white-noise $D_0$, obtaining a ``phase
chart'' of the regions of the parameters space where NDM and ANM
occur~\cite{CPSV17}. For large values of the noise $D_0\gg 1$, one
observes a simple monotonic behavior, as expected, because in that
case the effect of the underlying velocity field is negligible with
respect to the noise; by decreasing $D_0$, instead, one finds NDM for
small enough values of $\tau_S$, with a narrow region of ANM for $0.6
\lesssim \tau_S \lesssim 1$ (and $D_0\lesssim 5\times 10^{-3}$): in
particular, ANM is observed when the Stokes time of the particle and
the characteristic time of the velocity field are comparable, namely
for $\tau_S\sim \tau^*=1$.  All these peculiar phenomena can be traced
back to the general expression for the average of Eq.~(\ref{eq2})
\begin{equation}
\langle v_x\rangle=F\tau_S + \langle U_x(x,y)\rangle. 
\label{vx}
\end{equation}
This shows that the external force has a twofold effect: it drives the
tracer along its direction, but at the same time, it pushes the
particle towards specific regions of the underlying velocity field.
ANM and NDM, thus, emerge from the subtle competition between the
terms $F\tau_S$ and $\langle U_x(x,y)\rangle$.

\subsection{Negative differential mobility}
\begin{figure}[!tb]
\centering
\includegraphics[width=0.45\columnwidth,clip=true]{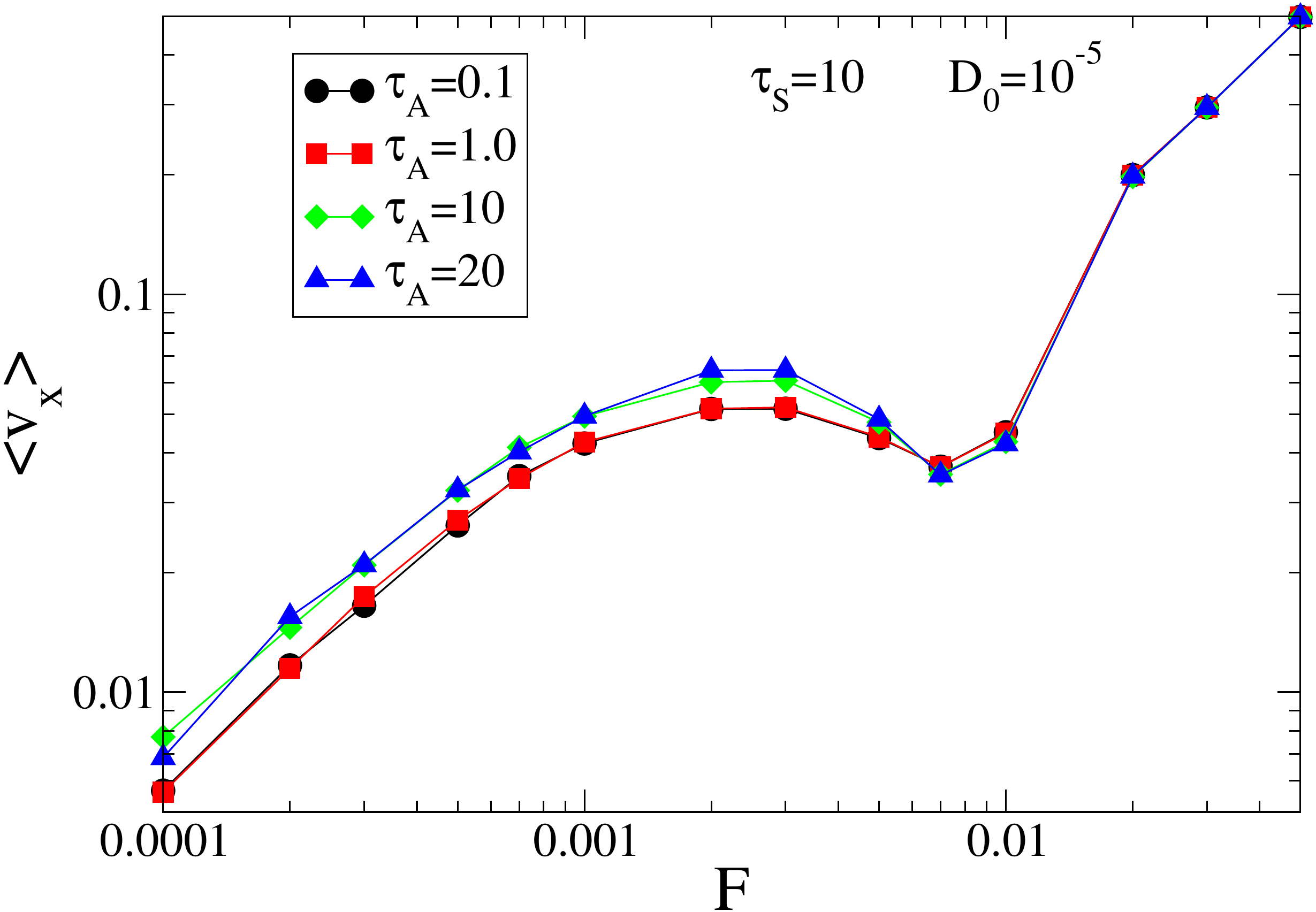}
\includegraphics[width=0.45\columnwidth,clip=true]{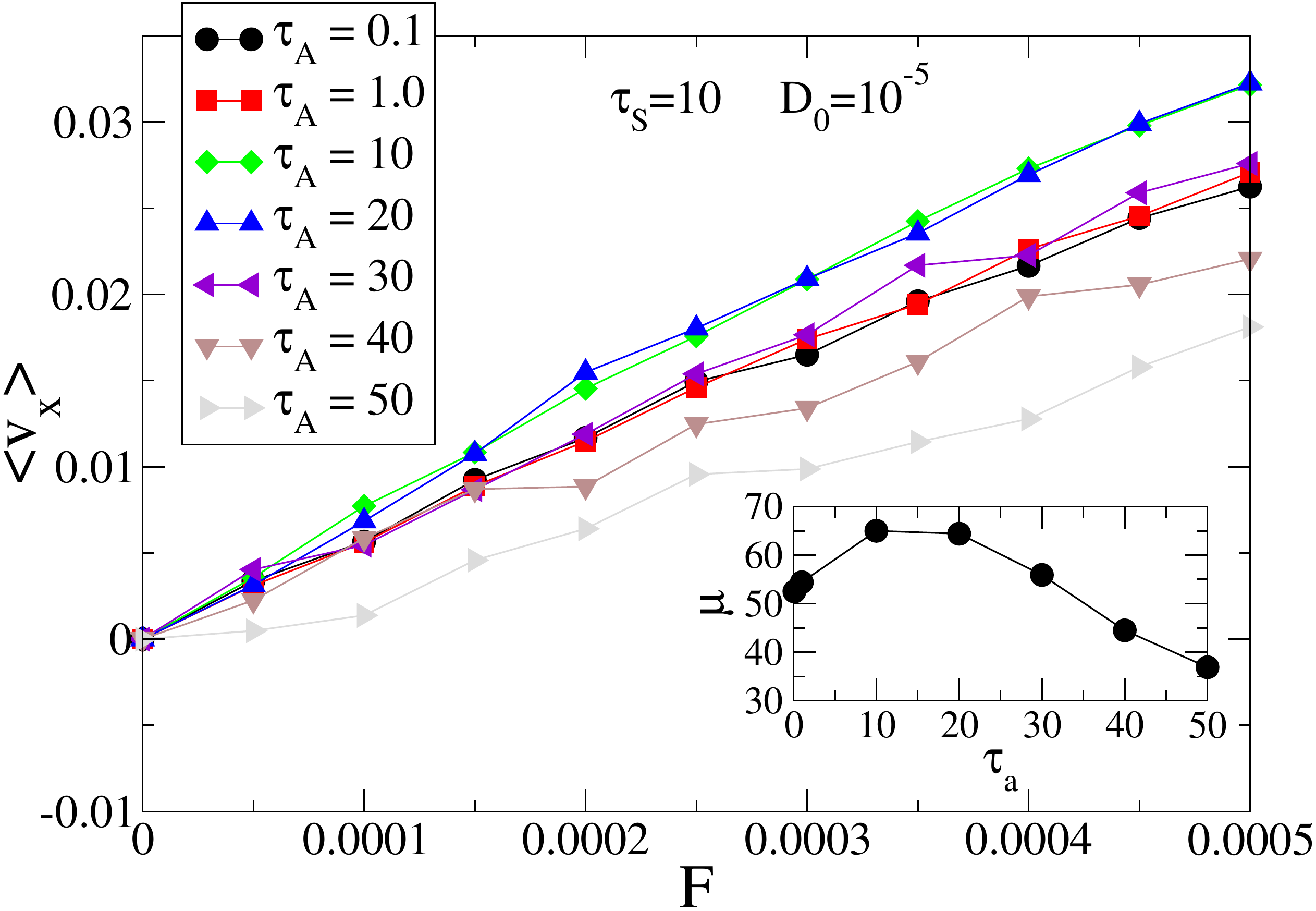}
\caption{Force-velocity relation $\langle v_x\rangle(F)$ for
  $\tau_S=10$, $D_0=10^{-5}$ and different values of $\tau_A$. Data
  are obtained averaging over $\sim 10^4$ initial conditions and noise
  realizations. Left: Note the non-monotonic behavior, corresponding
  to negative differential mobility. Right: Zoom in the small force (linear) regime,
  from which the mobility $\mu$ is obtained (inset).}
\label{fig1}
\end{figure}
In order to investigate the effect of colored noise, i.e. the role of
the new parameter $\tau_A$, we first study the case with Stokes time
$\tau_S=10$ and $D_0=10^{-5}$. The average velocity $\langle
v_x\rangle$ is reported in Fig.~\ref{fig1} as a function of the
applied force $F$. Here $\langle\cdots\rangle$ is computed over many
different ($\sim 10^4$) initial conditions and on long trajectories.
One observes first a linear increase of the velocity and then a
non-monotonic behavior.  Eventually, for larger values of the force,
the linear behavior is recovered, as expected.  The intermediate
region shows that even in the case of finite $\tau_A$, the phenomenon
of NDM occurs, suggesting that it is quite robust with respect to the
kind of noise considered in the model. In the right panel of
Fig.~\ref{fig1} we focus on the small force, linear regime to study
the mobility $\mu=\lim_{F\to 0}\langle v_x\rangle/F$, that turns out
to be a non-monotonic, weakly dependent function of the correlation
time $\tau_A$ (see inset).

\subsection{Absolute negative mobility}
\begin{figure}[!tb]
\centering
\includegraphics[width=0.45\columnwidth,clip=true]{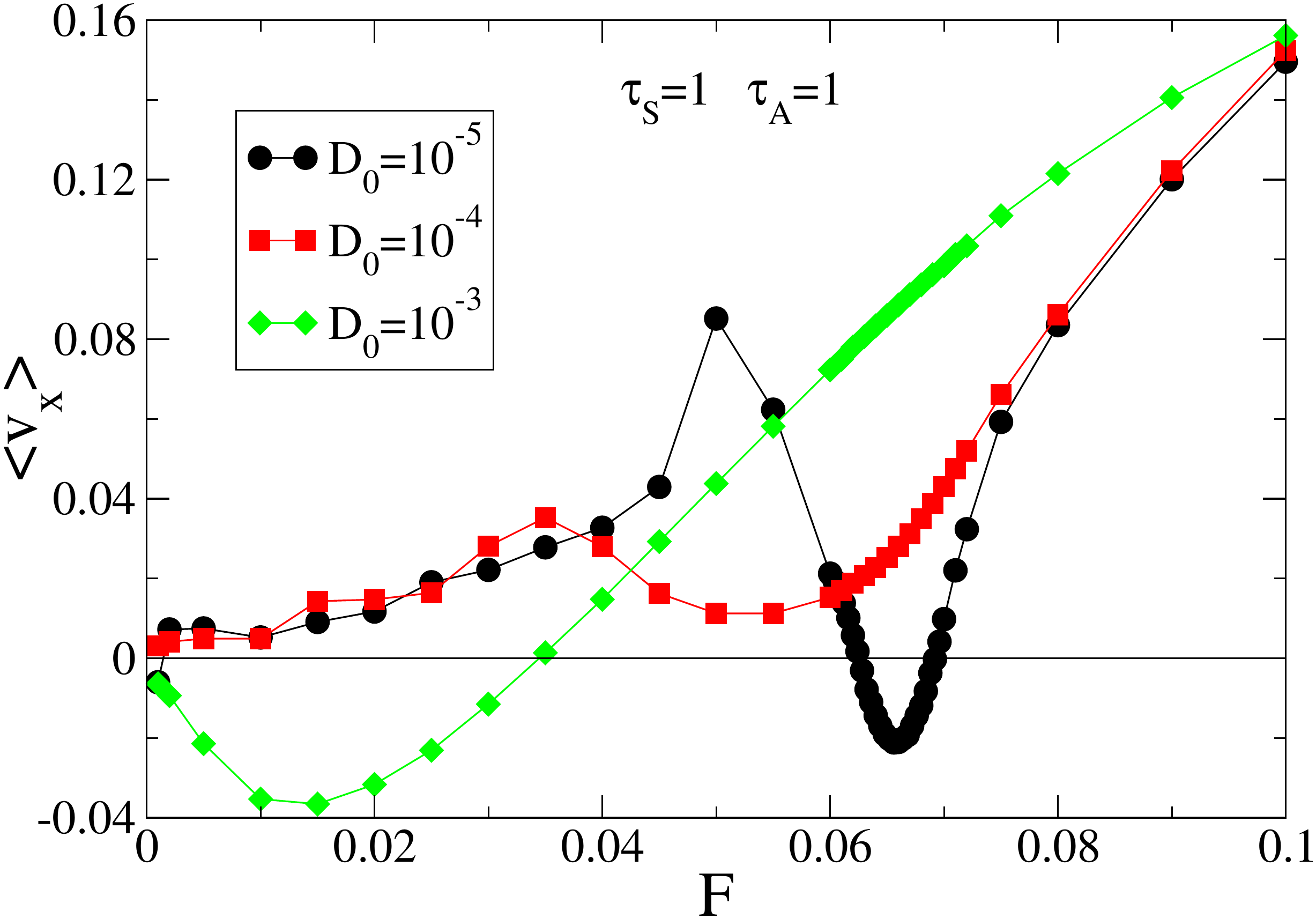}
\includegraphics[width=0.45\columnwidth,clip=true]{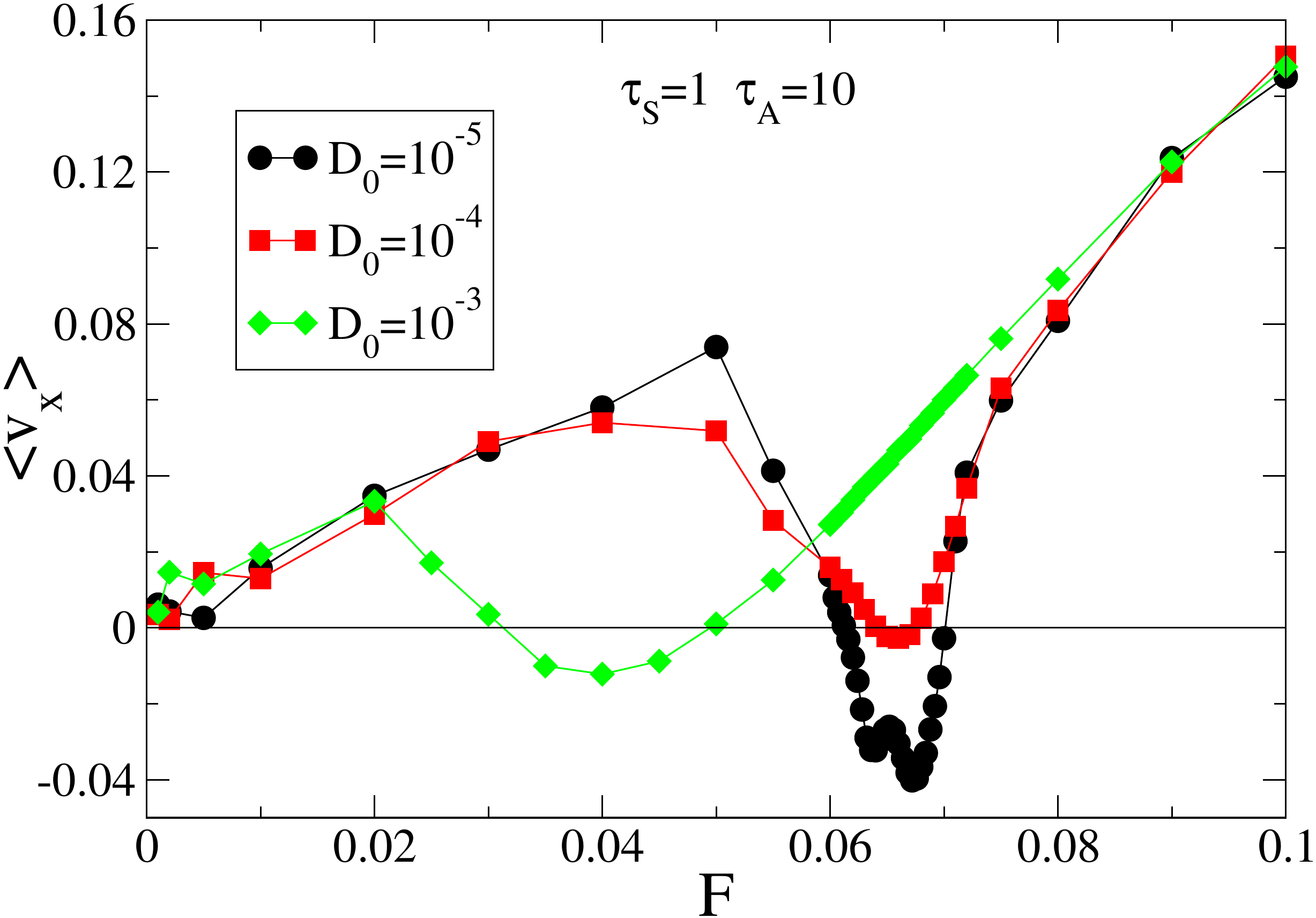}
\includegraphics[width=0.45\columnwidth,clip=true]{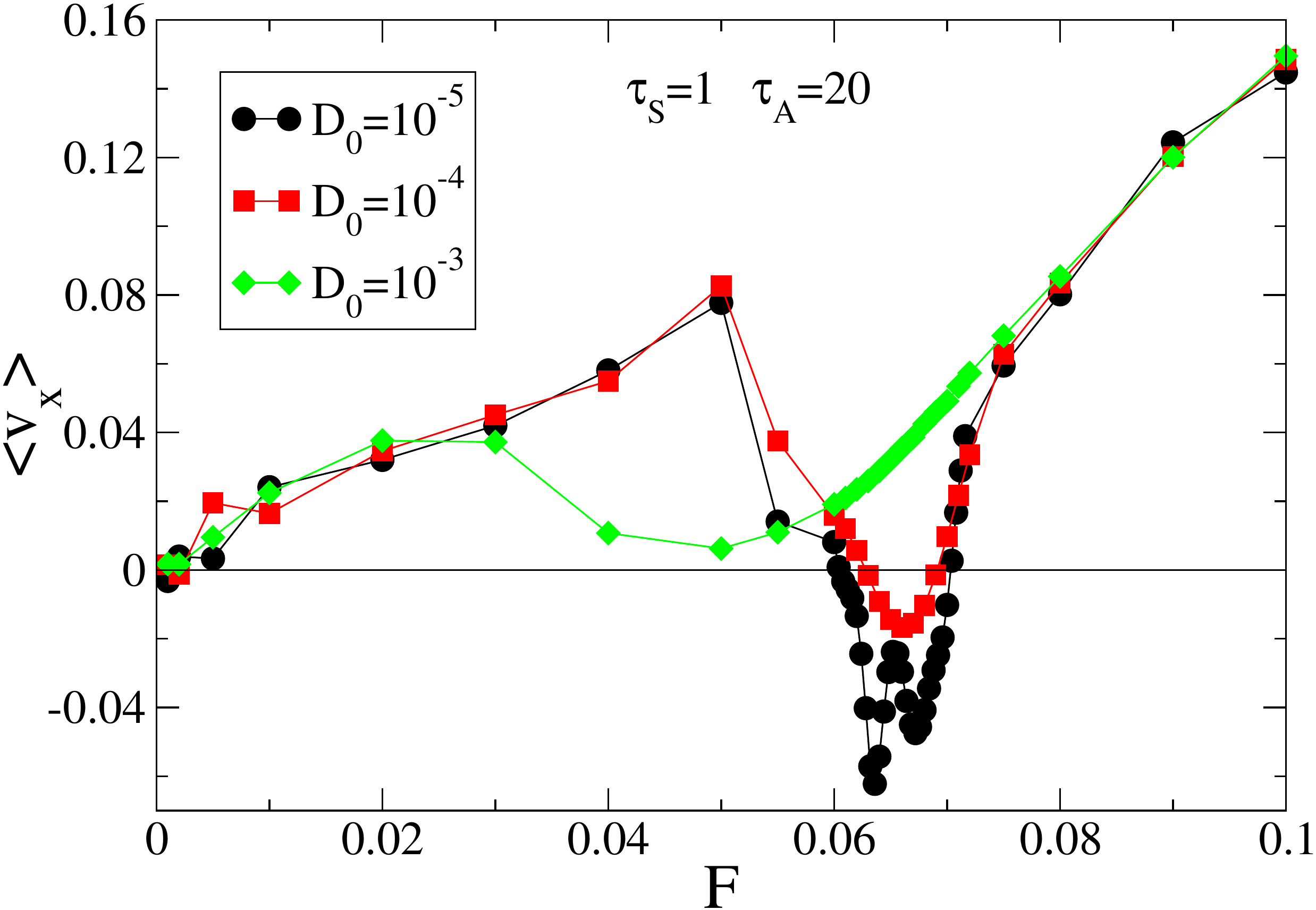}
\includegraphics[width=0.45\columnwidth,clip=true]{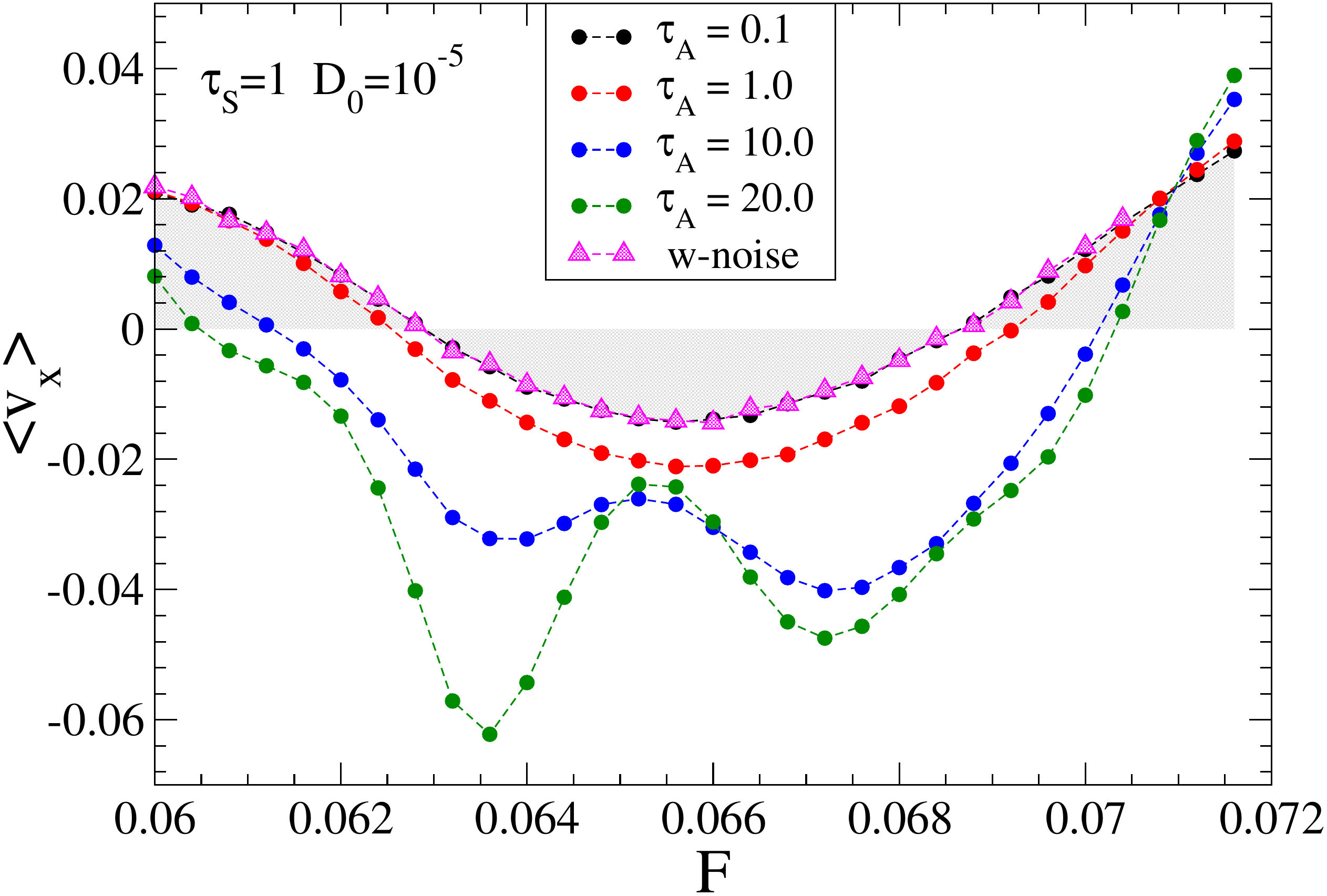}
\caption{Force-velocity relation $\langle v_x\rangle(F)$ for
  $\tau_S=1$, and several values of $D_0=10^{-5}$ and $\tau_A$. In
  some cases absolute negative mobility is observed.  Bottom right
  panel: zoom of a ANM region, showing a complex behavior
  characterized by two minima (the grey area represents the regions
  where the average velocity for the white noise case,
  $\tau_A=0$, changes sign).}
\label{fig2}
\end{figure}
A more interesting behavior is observed if the Stokes time and the
characteristic time of the velocity field are of the same order,
$\tau_S\sim \tau^*$. In this case, for the passive inertial tracer
($\tau_A=0$) it has been shown~\cite{SCPV16,CPSV17} that there exists
a region of the parameters space $(\tau_S,D_0)$, where absolute
negative mobility can be observed.  Here we investigate the robustness
of this phenomenon when a more general model, including colored noise,
is considered, and show how the phase diagram gets modified.

We first focus on the case with Stokes time $\tau_S=1$ (see
Fig.~\ref{fig2}).  We investigate the behaviors of the force-velocity
curve as a functions of the time $\tau_A$ and of the noise amplitude
$D_0$. For $\tau_A=1$ we find ANM for $D_0=10^{-5}$ and $D_0=10^{-3}$,
but not for $D_0=10^{-4}$. Interestingly, in the case $D_0=10^{-3}$,
we observe a negative \emph{linear} response (green diamonds): this
behavior does not violate any fundamental principle beacause the
system is out of equilibrium even when the external force is zero, due
to the non-gradient form of the velocity field $\mathbf U$.  For
$\tau_A=10$ we find ANM for all explored values of $D_0$; for
$\tau_A=20$ we find ANM for $D_0=10^{-5}$, and $D_0=10^{-4}$, but not
for $D_0=10^{-3}$. With respect to the passive case ($\tau_A=0$) we
find that the region of the parameter space where ANM is observed is
enlarged for finite $\tau_A$, see Table~\ref{tab1}.

\begin{table}[!tb]
\centering
\begin{tabular}{| c | c | c | c | c|}
  \hline			
 $\tau_S=1$ & $\tau_A=0$& $\tau_A=1.0$ & $\tau_A=10$  &  $\tau_A=20$ \\
\hline
 $D_0=10^{-3}$ & ANM & ANM & ANM & NDM \\
 $D_0=10^{-4}$ & NDM & NDM & ANM & ANM \\
 $D_0=10^{-5}$  & ST & ANM & ANM & ANM \\
  \hline  
\end{tabular}
\caption{Table showing the regions of the parameter space where ANM
  and NDM is observed, for $\tau_S=1$ (ST means standard,
  i.e. $d\langle v_x\rangle/dF>0$ in all the range of explored values
  of $F$).}
\label{tab1}
\end{table}

In the bottom right panel of Fig.~\ref{fig2} we focus on the case
$D_0=10^{-5}$ and zoom in the region of ANM: we observe that for
finite $\tau_A$ an interesting two-minima behavior of the force
velocity relation is observed. It is also worth noting that the
negative minima of the velocity depend on $\tau_A$ and increase (in
absolute value) with increasing $\tau_A$. This suggests that the
phenomenon of ANM can be amplified by considering a finite
$\tau_A$. 
\begin{figure}[!tb]
\centering
\includegraphics[width=0.45\columnwidth,clip=true]{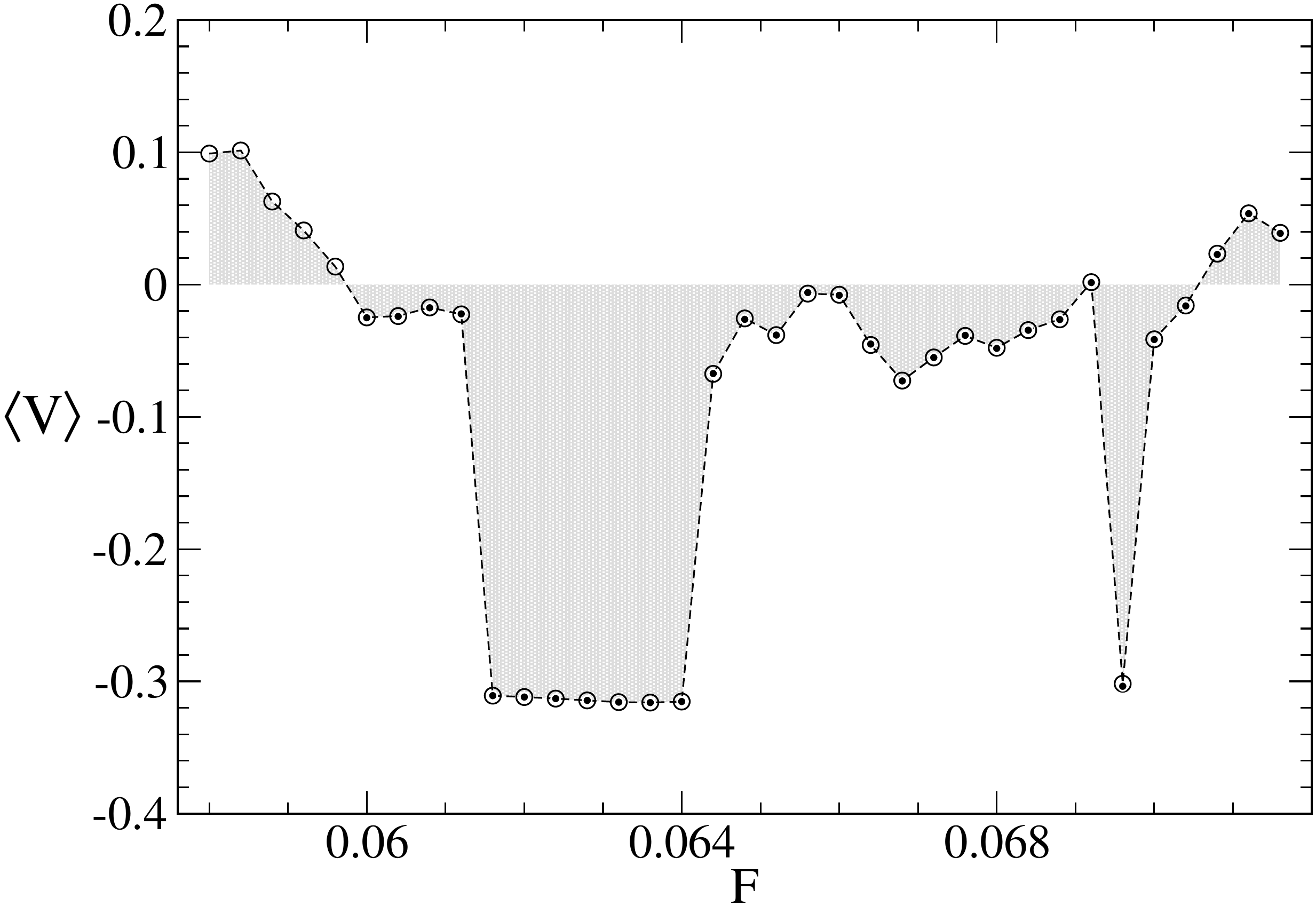}
\caption{Force-velocity relation for $\tau_S=1$ in the zero-noise 
($\tau_A\to \infty$) limit.\label{fig3}}
\end{figure}

\begin{figure}[!tb]
\centering
\includegraphics[width=0.45\columnwidth,clip=true]{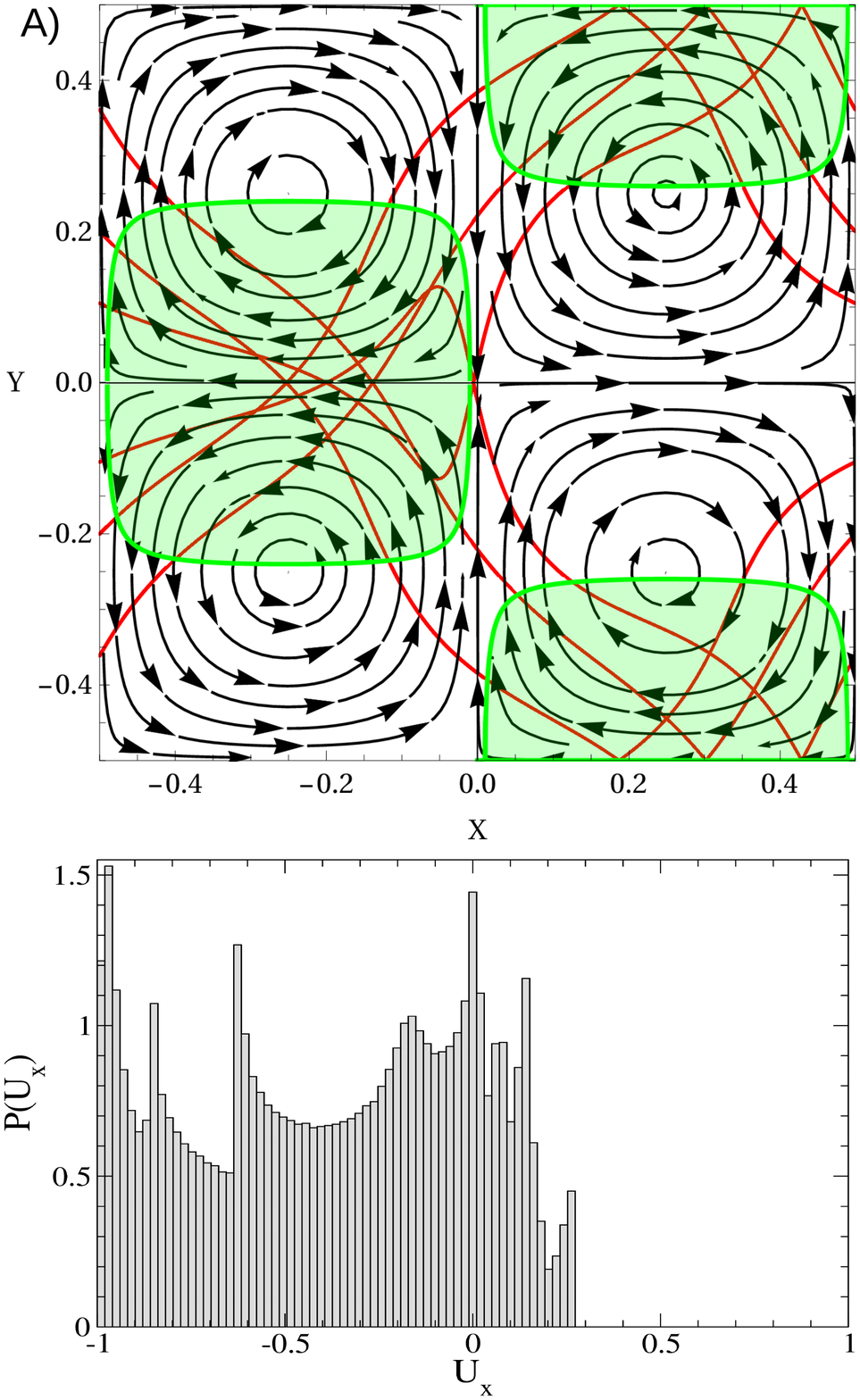}
\includegraphics[width=0.45\columnwidth,clip=true]{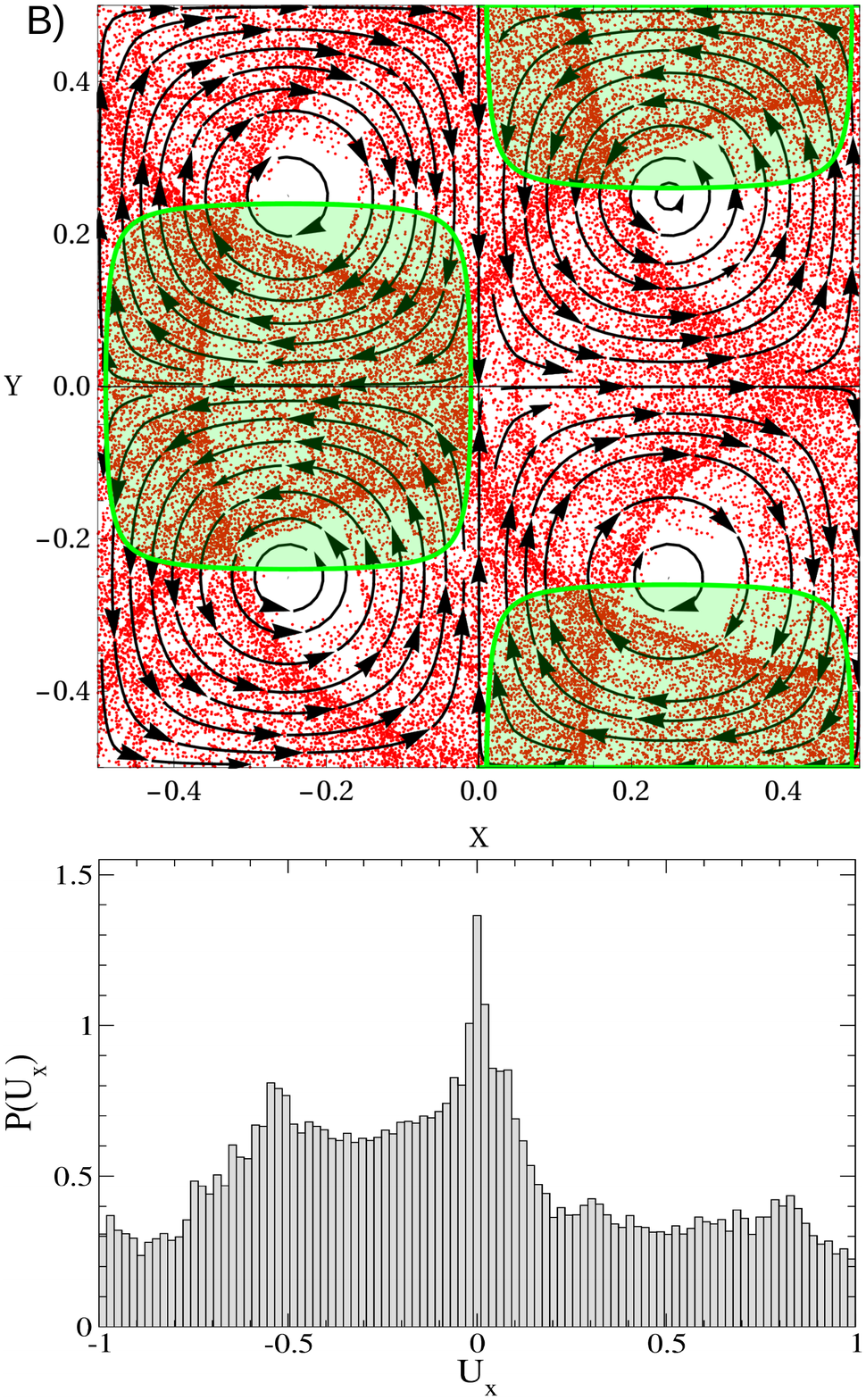}
\caption{Deterministic (noiseless) trajectory (red dots) sampled every
  $t_s=2000\Delta t$ and folded into the fundamental cell $C =
  [-L/2,L/2]\times[-L/2,L/2]$ for two cases: $\tau_S=1, F=0.063$ (left
  panel) and $\tau_S=1, F=0.065$ (right panel).  Left: the folded
  trajectory is periodic and mainly visits the regions of $C$ where
  $U_x(x,y) + F \tau_S \leq 0$ (shaded green areas). 
  Right: the folded trajectory looks chaotic but
  again points are denser in the regions where $U_x(x,y) + F\tau_S \leq 0$. 
  Bottom panels reports the normalized histograms of the values $U =
  U_x(x,y)$ collected along the trajectories showing the asymmetry
  with respect to the origin: $P(-|U|) > P(|U|)$.  This dynamical
  symmetry breaking is another signature of the presence of negative
  mobility states.\label{fig4}}
\end{figure}
More specifically, this seems to be related to the underlying
deterministic structure. Indeed, in the limit $\tau_A\to \infty$ the
stochastic terms $w_x$ and $w_y$ appearing in Eqs.~(\ref{eq2})
and~(\ref{eq2a}) become small (order $\sqrt{D_0/\tau_A}$) constants,
depending on the initial conditions. In Fig.~\ref{fig3} we show the
force-velocity relation for $\tau_S=1$ and zero noise, namely
$w_x=w_y=0$. In this deterministic case we find two deep negative
minima (note the scale of ordinates), suggesting that the underlying
deterministic dynamics governs the behavior observed in the presence
of noise. This explanation has been carefully described
in~\cite{SER07} for a one-dimensional system. 

Figure~\ref{fig4} reports $M=5 \times 10^4$ points sampled from a
deterministic (noiseless) trajectory of the particle with $\tau_S=1$
and re-folded into the fundamental cell
$C=[-L/2,L/2]\times[-L/2,L/2]$, for forces $F=0.063$ (left panel) and
$F=0.065$ (right panel). 
The figures indicate that the dynamics for
$F=0.063$ is regular/periodic while for $F=0.065$ it looks
``chaotic''; however, in both cases the trajectory spends more time in
the regions $R = \{(x,y)\in C\;|\;U_x(x,y)+ F\tau_S \leq 0\}$
marked by green-shaded domains. 
The extension and the contour of these regions is obtained by
solving the inequality for the explicit form of $U_x(x,y)$:
$$
\sin(kx) \cos(ky) \leq -\dfrac{F\tau_S}{U_0} \equiv -W.  
$$
The results are the two domains represented in Fig.~\ref{fig4}:
$$
D_1 = 
\begin{cases}
x\in[-L/2,0] \\
|y| \leq \dfrac{1}{k} \arccos\bigg[\dfrac{W}{|\sin(kx)|} \bigg]
\end{cases}
$$
and 
$$
D_2 = 
\begin{cases}
x\in[0,L/2] \\
|y| \geq \dfrac{1}{k} \arccos\bigg[\dfrac{-W}{|\sin(kx)|} \bigg].
\end{cases}
$$
For the values of $F$ for which ANM is observed, Fig.~\ref{fig3}, the   
dynamics preferentially occupies these domains so that the average (\ref{vx}) turns
to be certainly negative.

The bottom panels of Fig.~\ref{fig4} show the corresponding histograms of the set
of values $U_k = U_x[x(t_k),y(t_k)]$, $k=1,...,M$ evaluated over the
sampled points.  Both histograms exhibit a neat asymmetry with respect
to the origin, $P(-|U|)>P(|U|)$, indicating that in Eq.~(\ref{vx})
the average $\langle U_x(x,y)\rangle$ gets mainly negative
contributions and dominates over $F\tau_S$.

\begin{table}[!tb]
\centering
\begin{tabular}{| c | c | c | c | c|}
  \hline			
 $\tau_S=0.65$ & $\tau_A=0$& $\tau_A=1.0$ & $\tau_A=10$  &  $\tau_A=20$ \\
\hline
 $D_0=10^{-3}$ & ANM  & NDM & NDM & NDM \\
 $D_0=10^{-4}$ & NDM  & NDM & NDM & NDM \\
 $D_0=10^{-5}$ & ANM  & NDM & NDM & NDM \\
  \hline  
\end{tabular}
\caption{Table showing the regions of the parameter space where ANM
  and NDM is observed, for $\tau_S=0.65$.} 
\label{tab2}
\end{table}

We have performed the same analysis also for $\tau_S=0.65$:
Table~\ref{tab2} summarizes our results.  In particular, note that for
$\tau_S=0.65<\tau^*$ we don't observe ANM, but only NDM, at variance
with the white-noise case $\tau_A=0$, where ANM occurs for some values
of $D_0$.

\section{Conclusions}

In this paper, we have studied the nonlinear response to an external
force of an active particle, with persistence time $\tau_A$, in the
presence of a laminar flow. We have focused on the behavior of
the average velocity of the particle as a function of the applied
force, exploring a region of the model parameter space,
$(\tau_S,\tau_A,D_0)$, where $\tau_S$ is the Stokes time of the
particle and $D_0$ the amplitude of the active noise. We found that
the force-velocity relation of the particle can show non-monotonic
behaviors, including negative differential and absolute mobility.  Our
results indicate that the response of active matter in the presence of
an underlying velocity field can show counterintuitive phenomena, such
as NDM or ANM.  In particular, this can be applied for sorting and
selection of particles with different activity (defined via the active
time $\tau_A$), exploiting the different response to an external
force.

In general, anomalous response behavior can be expected in the nonlinear
response regime, or/and out of equilibrium, namely in the presence
of currents, when the time-reversal symmetry and detailed
balance are violated by the dynamics. This happens in our model, where, even for zero external force, 
the system does not satisfy
detailed balance, due to the non-gradient form of the velocity field
and due to the active nature of the tracer.

It could be also interesting the study of more complex situations
where several interacting active particles are considered, or
including the effect of the particle motion on the structure of the
surrounding fluid, or the presence of particular boundary
conditions. Moreover, the investigation of the diffusion properties in
the absence and in the presence of the external force requires further
analysis, in particular in the light of the non-equilibrium
fluctuation-dissipation relations, connecting mobility and diffusion
coefficient.

\section*{Bibliography}

\bibliographystyle{iopart-num}
\bibliography{biblio2}

\end{document}